# Copper Thin Film Deposition by An Indigenous Unbalanced Type DC Magnetron Sputtering System

SOUMIK KUMAR KUNDU[1], SAMIT KARMAKAR[1] and G. S. TAKI[1,*]

[1]Institute of Engineering & Management, Kolkata-700091, India
*Author for correspondence (gstaki@iemcal.com)

**Abstract.** Copper deposition has been carried out at various time span on glass slide and silicon substrate by using indigenously developed unbalanced type DC magnetron sputtering system. The main objective of this work is to study the crystalline structure of the deposited materials and also to calculate the crystallite grain size. As a transition metal, Copper nanoparticles and structures have several utilities in the field of photocatalytic and sensor applications. Such structures are utilized to provide free electrons that enhance optical and electrical properties of the photocatalytic sensor materials. These nano-catalysts enhance deposition rate and nucleation of graphitic Carbon Nitride – a popular photocatalyst. In this work, synthesized Copper thin film has been characterized by using X-Ray Fluorescence (XRF) and X-Ray Diffractometer (XRD).

## 1. Introduction

Magnetron Sputtering is one of the promising modern vacuum coating technique for the deposition of thin nano-metric layers of metals, non-metals and alloys. The sputtering phenomenon was first observed in the year of 1850, even though scientists started active research on this field in 1940. At the beginning, only diode sputtering was used. The magnetron sputtering concept was developed in mid-1970s. Since then this technique has been modified several times to attain the present state. Various researchers, found interest in this vacuum coating process for its several advantages like uniform deposition, less power requirements etc. [1,2]. N. Maréchal et al. used D.C. and R.F. magnetron sputtering for silver thin film deposition and studied its deposition rate [3]. J. Fu et al. from Applied Materials, Inc., California, deposited copper on a substrate using magnetron sputtering without any inert gas leak (self-sputtering process) and developed more uniform copper film than that synthesized at high pressure Ar leak in his system [4]. Kelly et al. has studied the magnetic field properties by changing its orientation from mirror to closed field and found a further coating uniformity in closed field unbalanced magnetron sputtering [5]. Magnetron sputtered uniform metal coating can be used for protecting materials surface. It has been found by some researchers that transition metal nitride coating is very much essential to protect mechanical tools [6]. Optimization of target to substrate distance and inert gas pressure are most essential criteria for uniform film thickness.

In magnetron sputtering, comparatively lower negative bias potential is required than the other sputtering method. This negatively biased target will act as a cathode and electrons circulate around the closed magnetic field lines created by two concentric magnetic poles. The inert gas molecules are ionized due to the collision with energetic electrons. This phenomenon causes a stable DC discharge creating energetic positive ions. The produced positive ions bombard the metal target to eject target atoms which will be eventually deposited on the substrate. Fig 1. shows the metallic and non-metallic sputter deposition scheme. The deposition rate may be varied by changing the substrate to target distance, applied bias potential to the target and working vacuum.

A DC unbalanced type magnetron sputtering setup has been indigenously designed and developed for metallic thin film deposition. Fig. 2 shows the indigenously designed & developed DC magnetron sputtering setup. This setup consists of a non-magnetic Tee shaped stainless-steel chamber of 100mm diameter with a substrate handling facility, three gas inlet ports along with vacuum gauges and a sputter-head. This sputter-head consists of a unbalanced magnet assembly and a target mounting facility. An unbalanced magnetic field is created by a concentric magnet assembly consisting of a coin shaped and an annular ring-shaped ferrite magnet. For the synthesis of uniform film, vacuum condition is an important factor which is obtained by using a double stage rotary vane pump. For sputtering, the target will be negatively biased by using a DC regulated power supply. Here in this work, copper thin film has been tried to deposit using this in-house DC magnetron sputtering setup. The synthesized film has been characterized by using XRF and XRD. From the result, the crystallinity of the deposited materials has been studied. The crystallite grain size is calculated from the XRD results by using Scherrer's equation:

$$D = k\lambda/\beta cos\theta \qquad (1)$$

Where k = 0.9 (Scherrer's constant for cubic crystal),
$\lambda$ = 1.5406 Å (wavelength of the X-Ray),
$\beta$ = FWHM (in radians) and
$\theta$ = Peak position (in radians).

This work is the prior work to the graphitic Carbon Nitride synthesis.

## 2. Experiment

In this magnetron sputter deposition experiment, a copper sheet of 50 mm diameter and 1 mm thickness has been utilized as a target material. Glass slides and Silicon wafer were used as a substrate here. Before the deposition, the air inside the chamber was evacuated using a double stage rotary vane vacuum pump and base vacuum pressure of $1 \times 10^{-2}$ mbar was achieved after 25 minutes of pumping down. After achieving base pressure, Argon gas has been injected into the chamber for creating inert atmosphere during deposition and a pressure of $9 \times 10^{-2}$ mbar was achieved. The pumping speed was reduced by throttling to attain an operating pressure of $1.1 \times 10^{-1}$ mbar and was continued throughout the experiment. The target was negatively biased keeping the substrate at ground potential during the experiment. The deposition was carried out at a stable 30 mA discharge current at -260 $V_{DC}$ for three different time spans i.e., 30 mins, 45 mins and 1 hour. Fig. 3 shows, the Sputter Deposition of Cu target materials over glass substrate. The deposited materials have been characterized by XRF and XRD.

## 3. Results

The deposited samples have been characterized using XRF (Jordon Valley EX 3600) and XRD (D8 Advanced Bruker AXS) at UGC-DAE CSR Kolkata center. The qualitative compositional analysis of copper deposited glass samples was carried out by using X-ray Fluorescence at 12 KV input voltage and an emission current of 50µA under the vacuum pressure of 540 mbar. The resulting data were plotted between the energy range of 1-10 KeV. Fig. 4(a) and 4(b) show the compositional analysis of bare glass slide and sputter deposited glass slide. A characteristic peak of copper is obtained at 8.24 KeV for sputter deposited sample.

The crystallinity of the deposited materials was characterized by XRD at 3º incidence angle. The XRD data plots of deposited materials on Si is shown in Fig. 5. In all the samples, a prominent peak of Cu [1 1 1] is obtained at 2$\theta$ = 43.26º (JCPDS Card No. 85-1326). Prominent peaks of $Cu_2O$ are also observed at 2$\theta$ angles of 29.42º, 37.17º and 62.2º for the crystallinity of [1 1 0], [1 1 1] and [2 2 0] respectively. The peaks show fairly good correlation ship according to JCPDS Card No. 03-0898. The crystallite grain sizes are calculated from the XRD results for abovementioned crystalline structures. The XRD result shows the average grain size of 11.983 nm for Cu [1 1 1] is obtained. The XRD results also show the average grain size of 19.064 nm, 5.238 nm and 2.409 nm for $Cu_2O$ [1 1 0], [1 1 1] and [2 2 0] respectively. So, the overall results show that, during copper deposition cuprous oxide ($Cu_2O$) is also produced.

## 4. Conclusion

As the XRF provides the qualitative compositional analysis data, quantitative composition of the deposited materials cannot be determined from it. From the XRD pattern it is clearly understood that not only pure copper is deposited but also cuprous oxide is formed and deposited during the experiment. Inadequate vacuum inside the sputtering chamber is the main reason behind it. It is also known that; pure copper has the tendency to get oxidized during the time of deposition. The oxidation probability can be further reduced by achieving high vacuum.

## Acknowledgements

The authors would like to show their extreme gratitude to The Institution of Engineers (India) for sanctioning a project fund under R&D Grant-in-Aid scheme R.5/2/PG/2018-19/RDPG2018004 to carry out the experimental work. The authors are thankful to Dr. Abhijit Saha, Director, Dr M. Sudarshan and Dr. P. V. Rajesh of UGC-DAE CSR, Kolkata Centre for carrying out the characterization work.

# Figures

Insert figures here

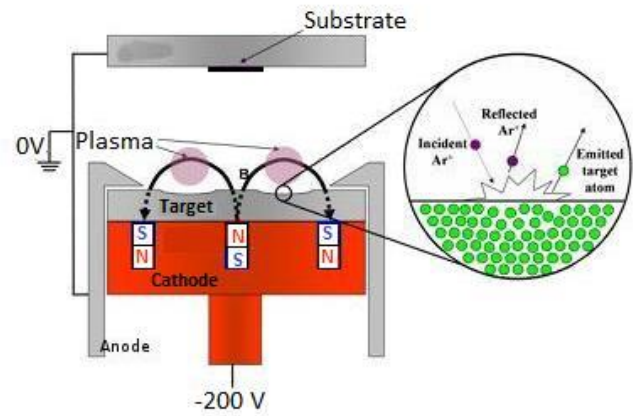

**Figure 1.**   The metallic and non-metallic sputter deposition scheme.

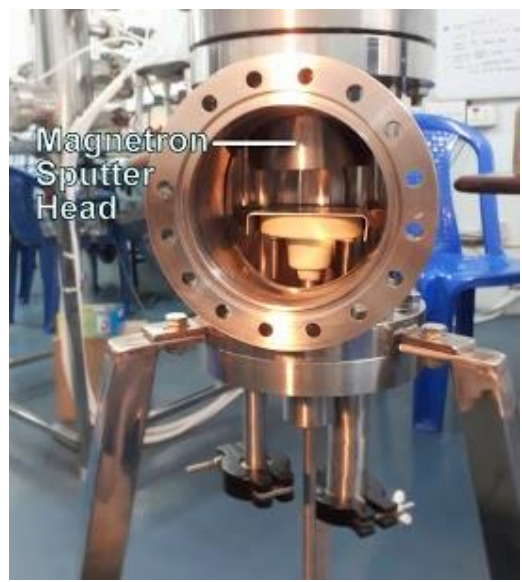

**Figure 2.** Indigenously designed & developed DC magnetron sputtering setup.

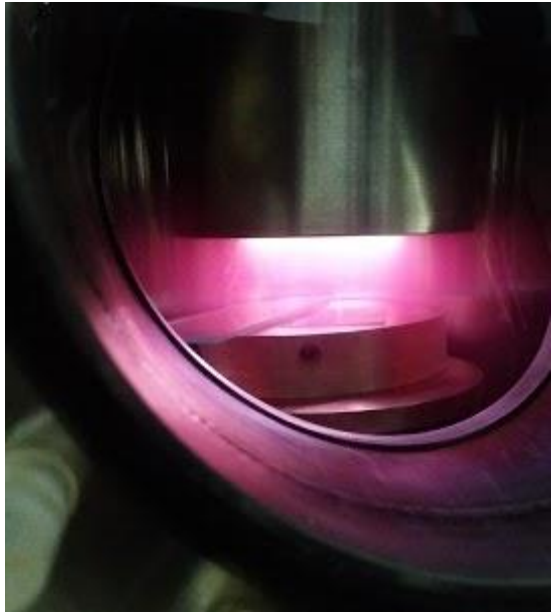

**Figure 3.** Sputter Deposition of copper sheet over glass substrate.

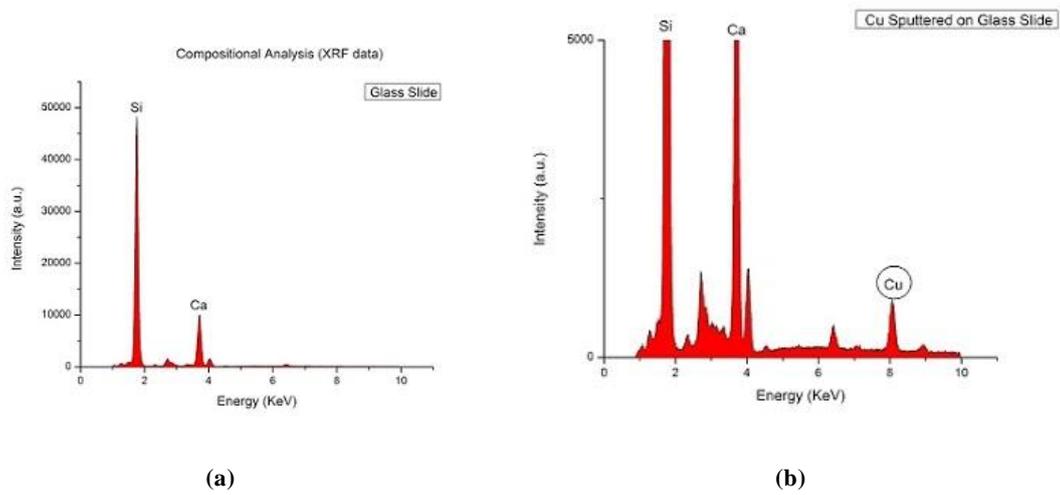

**(a)**          **(b)**

**Figure 4.** (a) compositional analysis of bare glass slide & (b) compositional analysis of the glass slide after sputter deposition.

6                                          *S K Kundu et al.*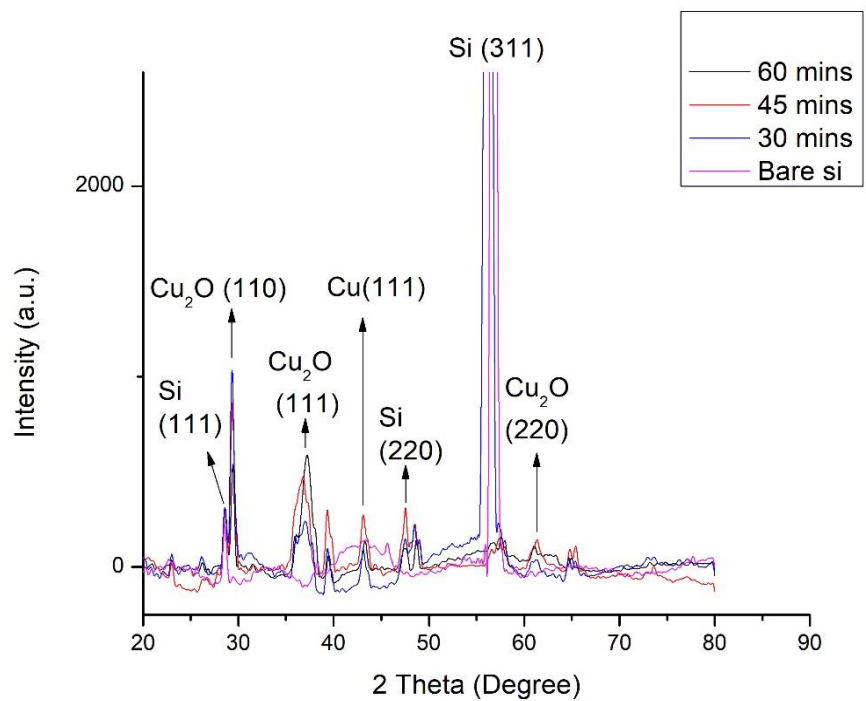

**Figure 5.** XRD pattern of the deposited thin film at three different times on Silicon substrate.